\documentclass[runningheads]{llncs}
\usepackage[T1]{fontenc}
\usepackage{graphicx,verbatim,tabularx,booktabs}
\usepackage{float}
\begin{document}
\title{When Repository Labels Are Not Image-Level Truth: A Supervision Auditing Framework for Chest Radiograph AI}
\titlerunning{A Supervision Auditing Framework for Chest Radiograph AI}

\author{Yesika Alexandra Agudelo-Londoño\inst{1,2} \and
Jhon Wilmer Pino-Rom\'an\inst{1} \and
Brahian Carrera Rodr\'iguez\inst{1,2} \and
Jos\'e Miguel Casta\~neda-Bedoya\inst{1,2} \and
Juan Pablo G\'omez-L\'opez\inst{1,2} \and
Aura C. Puche-Sarmiento\inst{1} \and
Niharika S. D'Souza\inst{3} \and
Juan Sebastian Osorio-Valencia\inst{1} \and
Jon E. Duque-Grajales\inst{1} \and
Jazm\'in Ximena Su\'arez-Revelo\inst{1} \and
Jorge Mario V\'elez-Arango\inst{1} \and
Gabriel Castrill\'on\inst{1,4}}
\authorrunning{Y. A. Agudelo-Londoño et al.}
\institute{Medical Imaging and AI in Healthcare Research Group, Biosciences Center, SURA, Medell\'{\i}n, Colombia \and
Universidad de Antioquia, Medell\'{\i}n, Colombia\\
\email{yesika.agudelo@udea.edu.co} \and
IBM Research, San Jose, CA, USA \and
Neuroradiology Institute, Universit\"atsklinikum Erlangen, Erlangen, Germany}
  
\maketitle              

\begin{abstract}
Public chest X-ray repositories are widely used to train medical AI systems, yet their labels are typically extracted from radiology reports rather than verified directly on images. As a result, repository labels are often treated as image-level ground truth without validating whether they reflect what is actually visible in the radiograph. We introduce \emph{Repository Supervision Auditing (RSA)}, a framework that evaluates repository-derived labels against expert image-level annotations before model development. Using cardiomegaly in MIMIC-CXR as a case study, RSA compares repository labels with radiologist-reviewed image annotations, characterizes disagreement sources, and builds a curated cohort for deployment-oriented evaluation. Repository-derived cardiomegaly labels showed near-zero agreement with expert image-level assessment, identifying only 1\% of expert-confirmed cases. Most discrepancies resulted from non-mention rather than explicit report negation, with expert-confirmed cardiomegaly identified in nearly half of studies assigned a repository-derived \emph{No Finding} label. Using the resulting expert-curated cohort, a DenseNet121 model achieved a test ROC-AUC of 0.853. These findings show that repository labels may not reliably represent image-level truth and highlight supervision auditing as a critical step for developing trustworthy medical imaging AI.
\end{abstract}

\section{Introduction}

Large-scale public chest radiograph repositories have become the foundation of modern AI for medical imaging. Datasets such as MIMIC-CXR, CheXpert, ChestX-ray14, and PadChest provide millions of radiographs paired with reports or automatically derived labels and now underpin the development, pretraining, and evaluation of deep learning and vision-language models for chest radiograph interpretation~\cite{johnson2019mimic,irvin2019chexpert,wang2017chestxray,bustos2020padchest,boecking2022making,tiu2022expert}. As these repositories increasingly support foundation models and deployment-oriented AI systems, the validity of the supervision used during model development has become increasingly important.

Most repository labels are not obtained through direct image review but are automatically extracted from radiology reports using natural language processing systems such as CheXpert and NegBio~\cite{irvin2019chexpert,peng2018negbio}. Although this approach enables scalable dataset construction, report-derived labels reflect clinical documentation rather than a comprehensive image-level assessment.  Stable, morphologic, or clinically secondary findings may be visually evident yet omitted from reports, causing repository labels to encode reporting practice as much as radiographic truth. Previous work has improved report-label extraction~\cite{peng2018negbio,irvin2019chexpert}, characterized hidden stratification~\cite{oakden2020hidden}, documented discrepancies between report-derived and image-level labels ~\cite{jain2021visualchexbert}. More recent work audits the reliability of automated label extraction itself, showing report-to-label extractors can be unreliable and improved by large language models~\cite{stock2026evaluating}; RSA is complementary, auditing a distinct level of validity—whether a correctly extracted report label is an appropriate image-level target for a visually defined finding. However, repository-derived labels are still routinely treated as image-level ground truth during model development. As AI systems trained on public repositories move toward clinical deployment, this assumption becomes increasingly consequential because model performance ultimately depends on the validity of the supervision itself. Supervision should therefore be audited before, rather than after, model development.

We introduce \emph{repository supervision auditing} (RSA), a framework that treats the supervisory signal itself as the object of evaluation. Rather than proposing a new model architecture, RSA evaluates whether repository-derived labels constitute valid image-level learning targets. Using cardiomegaly in MIMIC-CXR as a case study, we find near-zero agreement between repository and expert image-level labels, driven predominantly by non-mention rather than explicit negation, and construct a balanced, quality-controlled cohort for deployment-oriented evaluation.

\section{Repository Supervision Auditing}
\begin{figure}[t!]
\centering
\includegraphics[width=\textwidth]{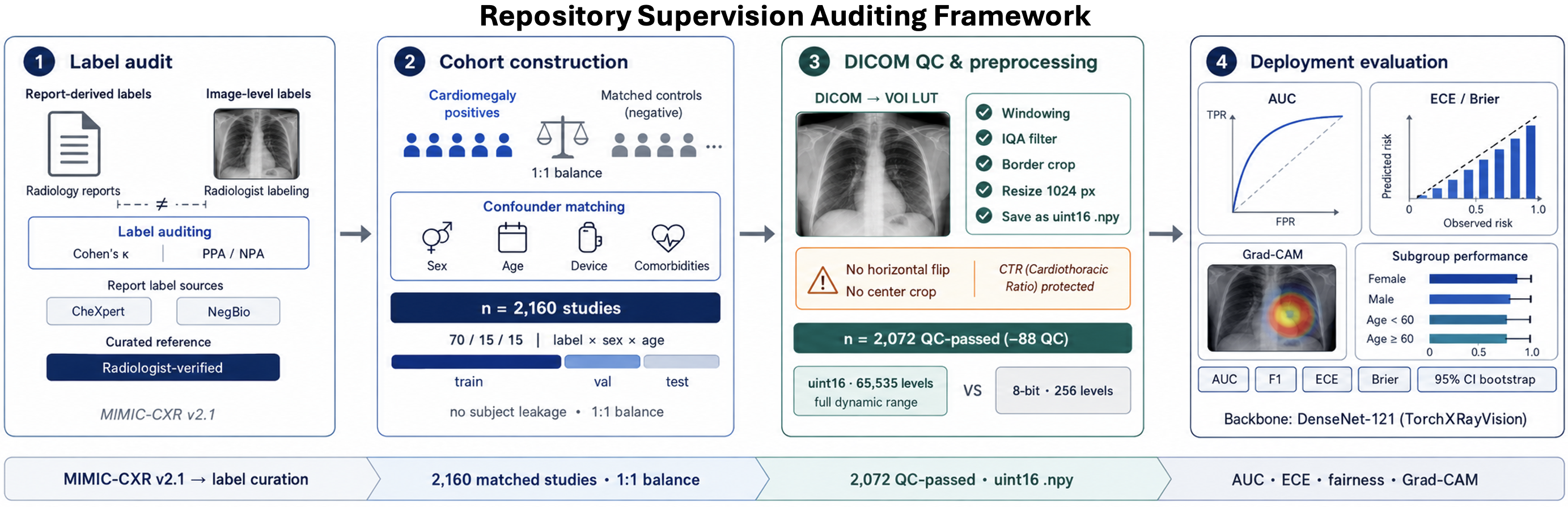}
\caption{Overview of the proposed \emph{repository supervision auditing} framework. Repository-derived labels are validated against expert image-level annotations to establish a curated reference standard, which is used to construct a balanced, quality-controlled cohort for deployment-oriented model development and evaluation.}
\label{fig:pipeline}
\end{figure}

The RSA framework evaluates whether repository-derived labels are suitable learning targets before model development, comprising four stages: (1) construction of an expert-curated image-level reference, (2) auditing repository derived labels against the expert reference, (3) construction of an expert-curated cohort for downstream model development, and (4) deployment-oriented model evaluation (see Figure~\ref{fig:pipeline}).

\medskip
\noindent\textbf{Expert-curated image-level reference.}
We retrospectively analyzed 2,160 unique subject-session studies from MIMIC-CXR~\cite{johnson2019mimic}, restricted to non-portable PA radiographs from inpatient admissions without ICU transfer. Board-certified radiologists independently reviewed each DICOM study using all available radiographic projections, including lateral views where present, together with the report, annotating findings through a custom web-based interface, with terminology standardized following RadLex, SNOMED CT, and Fleischner Society recommendations~\cite{Rubin2008Creating,Benson2020SNOMED,Bankier2024Fleischner}. Expert annotations were compared with CheXpert and NegBio labels from MIMIC-CXR-JPG~\cite{irvin2019chexpert,peng2018negbio}. Repository labels were binarized using an uncertainty-as-positive rule, mapping positive/\emph{uncertain} labels to \emph{positive} and negative/\emph{not mentioned} labels to \emph{negative}. Using the expert reading as reference, we computed Cohen's $\kappa$, positive percent agreement (PPA; sensitivity), negative percent agreement 
(NPA; specificity), positive predictive value (PPV), and negative predictive value (NPV). Cardiomegaly was selected as the index finding because it depends primarily on image-level cardiac-silhouette morphology; pleural effusion and atelectasis were evaluated as secondary findings.

\medskip
\noindent\textbf{Characterization of label discordance.}
Discordant cardiomegaly cases were reviewed at the report level and categorized as arising from (1) explicit negation, (2) assignment to the repository \emph{No Finding} label, or (3) non-mention despite other documented abnormalities, distinguishing report interpretation (explicit negation) from routine reporting practice (\emph{No Finding} or omission).

\medskip
\noindent\textbf{Expert-curated cohort construction.}
We constructed the downstream cohort using expert image-level annotations rather than repository-derived labels. From 14,200 relabeled examinations, all cardiomegaly-positive studies were selected and matched to negatives on ten demographic and imaging covariates (including age, sex, device category, and age-related thoracic findings). Matching reduced the largest standardized mean difference from 0.87 to 0.005 and achieved balance across all matching covariates ($|\mathrm{SMD}|<0.10$). We performed quality control directly on DICOM images, which included LUT transformation, VOI windowing, photometric normalization, polarity correction, border cropping, and resizing of the longest image dimension to 1,024 pixels while preserving 16-bit pixel precision~\cite{dapamede2025dicom}. Quality control excluded non-frontal examinations, annotated images, and studies with inadequate exposure, yielding a final cohort of 2,072 studies (1,020 positive and 1,052 negative). The curated cohort was split into subject-independent training ($n=1,453$), validation ($n=311$), and held-out test ($n=308$) sets using stratified sampling over label, sex, and age group to preserve class balance while preventing patient-level information leakage.

\medskip
\noindent\textbf{Deployment-oriented evaluation.}
We fine-tuned a DenseNet121 backbone from TorchXRayVision~\cite{cohen2022torchxrayvision}, initialized from the multi-dataset \texttt{densenet121-\allowbreak res224-\allowbreak all} weights (pretrained across seven public chest-radiograph datasets: NIH ChestX-ray14, PadChest, CheXpert, MIMIC-CXR, Google, OpenI, and RSNA), which provide a broad multi-pathology representation rather than a cardiomegaly-specific initialization. All layers were fully fine-tuned end-to-end on the expert image-level labels (not head-only), with the classification head replaced for the binary task, adapting this general representation to the audited target. Images were resized to the network's native input resolution while preserving the cardiothoracic ratio, and horizontal flipping was disabled because it reverses cardiac laterality.

Operating thresholds were selected exclusively on the validation cohort and fixed for all subsequent analyses. To validate the expert-audited cohort under deployment-oriented conditions~\cite{maierhein2024metrics}, we evaluated discrimination (ROC-AUC and PR-AUC), threshold-dependent performance (sensitivity, specificity, precision, and F1-score), calibration (Brier score and expected calibration error), subgroup performance across age and sex, robustness across clinically relevant evaluation cohorts, and qualitative Grad-CAM visualizations~\cite{selvaraju2020grad}. Unlike conventional benchmark evaluation, this protocol combines predictive performance, calibration, robustness, fairness, and explainability to provide a comprehensive assessment of clinical readiness. Confidence intervals were estimated using a class-stratified bootstrap with 1,000 resamples.

\section{Repository Supervision Audit Findings}

\begin{table}[b]
\caption{Agreement between repository-derived labels and expert image-level annotations on the matched MIMIC-CXR audit cohort ($N=2{,}160$). Results for cardiomegaly were identical across CheXpert, NegBio, and their union. PPA: positive percent agreement; NPA: negative percent agreement; PPV: positive predictive value.}
\label{tab:discordance}
\centering
\small
\begin{tabular}{lcccc}
\toprule
Finding & Cohen's $\kappa$ [95\% CI] & PPA & NPA & PPV \\
\midrule
Cardiomegaly (index) & 0.011 [0.004, 0.019] & 1.3\% & 99.8\% & 0.875 \\
Pleural effusion & 0.319 [0.190, 0.447] & 21.7\% & 99.7\% & 0.682 \\
Atelectasis & 0.120 [0.067, 0.171] & 7.9\% & 99.5\% & 0.655 \\
\bottomrule
\end{tabular}
\end{table}

\begin{table}[t]
\caption{Characteristics of the expert-curated cohort after matching and DICOM quality control ($N=2{,}072$). Matching achieved covariate balance across all ten variables ($|\mathrm{SMD}|<0.10$), reducing the largest imbalance (age) from 0.87 to 0.005.}
\label{tab:cohort}
\centering
\small
\begin{tabular}{lccc}
\toprule
Characteristic & Cardiomegaly & No Cardiomegaly & Overall \\
\midrule
Studies, $n$ & 1,020 & 1,052 & 2,072 \\
Age (years) & $64.8 \pm 15.2$ & $65.1 \pm 15.3$ & $65.0 \pm 15.2$ \\
Female, $n$ (\%) & 629 (61.7) & 655 (62.3) & 1,284 (62.0) \\
Cardiac device, $n$ (\%) & 103 (10.1) & 101 (9.6) & 204 (9.8) \\
\bottomrule
\end{tabular}
\end{table}

\noindent\textbf{Repository labels fail to recover image-level cardiomegaly.}
We first evaluated whether repository-derived labels agreed with expert image-level annotations. Repository-derived cardiomegaly labels showed near-zero agreement with expert image review ($\kappa$=0.011). Only 14 of 1,080 expert-confirmed cardiomegaly studies were labeled positive by the repository, yielding a PPA of 1.3\% despite a PPV of 0.875; CheXpert, NegBio, and their union produced identical results. In contrast, pleural effusion and atelectasis demonstrated higher agreement, indicating that repository label reliability is finding-dependent rather than uniformly poor (Table~\ref{tab:discordance}).

\medskip
\noindent\textbf{Discordance reflects non-mention, not explicit negation.}
We next examined the origin of discordant cardiomegaly labels. Of 1{,}066 false-negative repository
labels, only 55 (5.2\%) reflected explicit report negation; the remaining 1{,}011 (94.8\%) arose from
omission---979 (96.8\%) under a repository \emph{No~Finding} label and 32 in reports that documented
other abnormalities but not cardiomegaly. Indeed, expert review identified cardiomegaly in nearly half of
the studies labeled \emph{No~Finding} (979 of 1{,}998). Disagreement therefore reflects routine reporting
practice rather than label-extraction errors.

\begin{figure}[t]
\centering
\includegraphics[width=\textwidth]{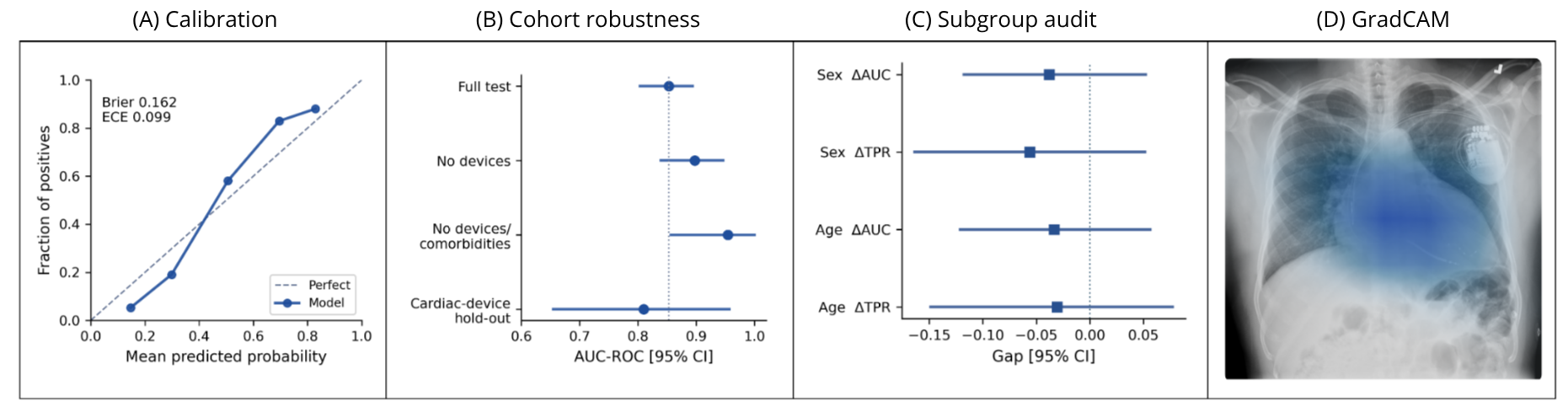}
\caption{Deployment-oriented evaluation using the expert-curated cohort.
(A) Reliability diagram showing predicted versus observed probabilities (Brier score = 0.162, ECE = 0.099).
(B) Robustness analysis across clinically relevant evaluation cohorts.
(C) Fairness analysis reporting differences in discrimination ($\Delta$AUC) and sensitivity ($\Delta$TPR) across age and sex; error bars denote 95\% confidence intervals.
(D) Representative Grad-CAM visualization suggesting that model attention concentrates on the cardiac silhouette rather than implanted cardiac devices.
}
\label{fig:eval}
\end{figure}

\begin{table}[t]
\caption{Downstream performance on the held-out test cohort ($N=308$). The operating threshold (0.436) was selected on the validation set using Youden's index. Confidence intervals were estimated using 1,000 class-stratified bootstrap resamples.}
\label{tab:perf}
\centering
\small
\begin{tabular}{lclc}
\toprule
Metric & Value [95\% CI] & Operating point & Value \\
\midrule
ROC-AUC & 0.853 [0.803, 0.893] & Sensitivity & 0.862 \\
PR-AUC & 0.824 [0.769, 0.878] & Specificity & 0.705 \\
Brier score & 0.162 & Precision & 0.740 \\
ECE (5 bins) & 0.099 [0.062, 0.138] & F1-score & 0.796 \\
\bottomrule
\end{tabular}
\end{table}

\medskip
\noindent\textbf{RSA enables construction of a trustworthy cohort.}
Guided by the audit findings, we constructed a balanced expert-curated cohort comprising 2,072 studies. Matching reduced the largest standardized mean difference from 0.87 to 0.005 and achieved balance across all matching covariates ($|\mathrm{SMD}|<0.10$). In addition, DICOM quality control further eliminated confounding factors related to the acquisition (Table~\ref{tab:cohort}).

\medskip
\noindent\textbf{Expert image-level supervision supports trustworthy downstream learning.}
Figure~\ref{fig:eval} illustrates the downstream evaluation (Rightmost block).  
Table~\ref{tab:perf} summarizes the downstream performance. Using the curated cohort, DenseNet121 achieved an ROC-AUC of 0.853 (95\% CI 0.803--0.893) and a PR-AUC of 0.824 (95\% CI 0.769--0.878) on the held-out test cohort. The model demonstrated moderate calibration (Brier score = 0.162, ECE = 0.099), maintained consistent discrimination across progressively cleaner evaluation cohorts, and showed no statistically significant differences in discrimination ($\Delta$AUC) or sensitivity ($\Delta$TPR) across age and sex. Grad-CAM visualizations concentrated model attention primarily on the cardiac silhouette rather than implanted cardiac devices, suggesting that model predictions were driven primarily by cardiac-silhouette morphology rather than device-related shortcuts (See example in Figure~\ref{fig:eval}). We interpret this as evidence that the expert-defined visual phenotype recovered through RSA is learnable within the target population, rather than as a comparative claim of superiority over report-derived supervision (see Limitations).

\begin{figure}[!ht]
\centering
\includegraphics[width=1\textwidth]{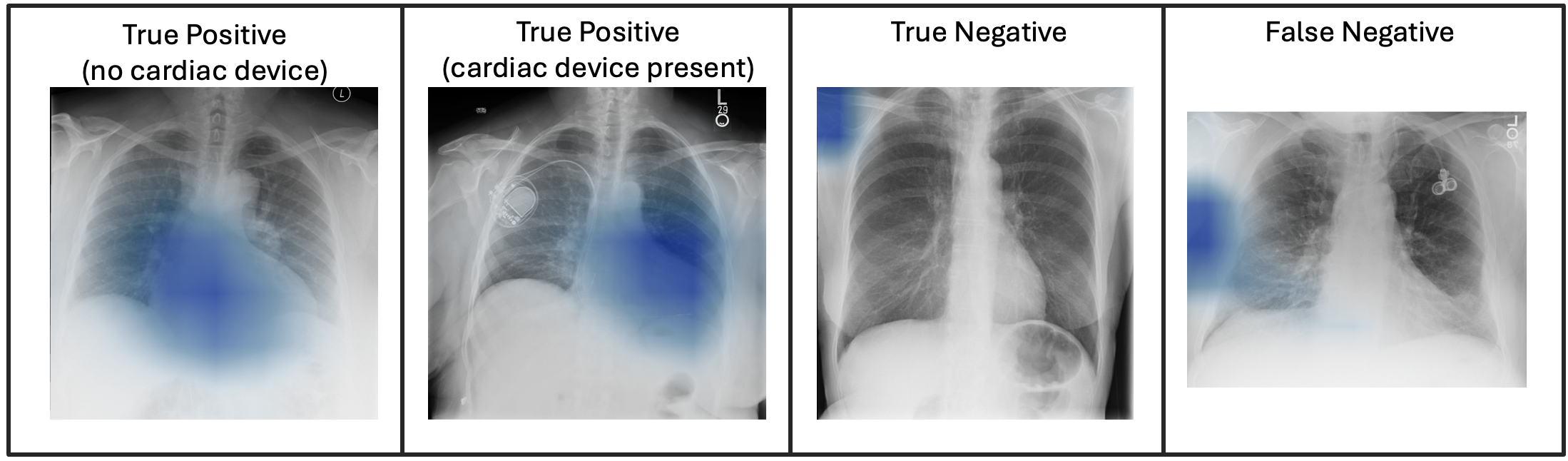}
\caption{
Representative Grad-CAM visualizations across diverse clinical presentations. (A) True positive without cardiac hardware: attention localizes to the enlarged cardiac silhouette. (B) True positive with an implanted pacemaker/defibrillator: attention remains focused on the enlarged silhouette rather than the device. (C) True negative: activation is diffuse and non-focal despite other thoracic findings in the absence of cardiomegaly. (D) False negative: attention is displaced from the cardiac silhouette; expert review identified bibasilar atelectasis and reduced lung volumes, which alter the apparent silhouette and provide a plausible explanation for the missed prediction.
}
\label{fig:gradcam}
\end{figure}

\medskip
\noindent\textbf{Qualitative model interpretation.}
To assess whether these observations generalized beyond individual examples, we examined Grad-CAM across diverse clinical presentations (Fig.~\ref{fig:gradcam}). In true-positive studies, attention consistently localized to the enlarged cardiac silhouette across both mild and marked cardiomegaly and remained anatomically focused in the presence of implanted cardiac devices. In the true-negative example, activation was diffuse and non-focal despite other thoracic findings, whereas the false-negative case exhibited displaced attention in the presence of bibasilar atelectasis and reduced lung volumes, which alter the apparent cardiac silhouette. Collectively, these observations suggest that the model learned features associated with cardiac enlargement rather than merely attending to the cardiac region, providing qualitative evidence of anatomically plausible feature learning. As Grad-CAM is sensitive to example selection, these visualizations are interpreted as qualitative, hypothesis-generating evidence rather than causal feature attribution.

\section{Discussion: From Repository Labels to Trustworthy AI}

This work introduces \emph{Repository Supervision Auditing (RSA)}, a framework for validating repository-derived supervisory signals before model development. Applied to MIMIC-CXR, RSA revealed that report-derived labels for cardiomegaly exhibit near-zero agreement with expert image-level assessment despite their widespread use for supervised learning. In contrast, pleural effusion and atelectasis showed substantially higher agreement, indicating that the reliability of repository-derived learning targets is finding-dependent rather than uniform.

RSA further showed that this disagreement stems primarily from routine reporting practice rather than report extraction errors. Most expert-confirmed cardiomegaly studies assigned a negative repository label were not explicitly negated, but absent from the report. This omission pattern is consistent with the acute-care setting of the source data, where reports may prioritize the presenting clinical question over stable or chronic findings. Consequently, repository-derived labels reflect clinical documentation rather than a complete inventory of radiographic abnormalities, and using them as image-level learning targets may introduce systematic supervision error for visually defined findings such as cardiomegaly.

Beyond identifying label discordance, RSA provides a practical workflow for developing deployment-oriented datasets. Expert image-level auditing, matched cohort construction, DICOM-level quality control, and reproducible evaluation produced a balanced reference cohort that supported reliable downstream learning using a standard DenseNet121 backbone. The resulting model demonstrated good discrimination, moderate calibration, no significant subgroup disparity, and robustness across clinically relevant evaluation cohorts. Qualitative Grad-CAM analyses further suggested that the model learned anatomically plausible representations by consistently localizing attention to the enlarged cardiac silhouette rather than imaging artifacts or implanted cardiac devices. Although these findings support the biological plausibility of the learned features, explainability alone does not establish clinical readiness, which requires external and prospective validation.

More broadly, our results suggest that RSA fills a missing stage in the medical AI development pipeline by validating repository-derived supervisory signals \emph{before} model development. Rather than replacing advances in model architecture, optimization, or foundation model pretraining, RSA complements them by ensuring that the supervisory signal itself is appropriate for the intended task. The matched-cohort construction and DICOM quality control used here are established best practices; they serve instrumentally to enable a controlled audit and are not themselves the contribution. RSA's contribution lies in treating the supervisory signal as the object of evaluation—what is audited, when, and how the findings inform subsequent development. 

\medskip
\noindent\textbf{Limitations and Future Work.}
This study has several limitations. First, the analysis was retrospective and based on a curated subset of a single public repository (MIMIC-CXR), restricted to non-portable inpatient radiographs from encounters without ICU transfer. Second, although the expert reference spans 52 findings, we focus on cardiomegaly with pleural effusion and atelectasis as representative comparisons; a comprehensive per-finding audit remains future work. Third, downstream evaluation used a single subject-independent train/validation/test split without external or prospective validation, and the modest test cohort limited power for subgroup and calibration analyses. Fourth, we do not present a head-to-head comparison between models trained on repository-derived versus expert image-level labels: within the audited cohort, repository labels identify almost no positive cases (PPA 1.3\%), leaving no positive-class support, and a separately constructed report-positive cohort would differ systematically in acquisition and case mix, confounding supervision source with these factors. Quantifying the practical benefit of RSA under adequate label overlap is left to future work.

Future work will extend RSA across additional repositories, imaging modalities, and disease phenotypes to determine where repository-derived learning targets are sufficient and where expert auditing provides the greatest benefit. External and prospective validation will establish the generalizability of RSA and quantify the independent contributions of expert supervision, cohort curation, and image quality control to downstream model performance.

\section{Conclusion}
Repository-derived labels should not be assumed to represent image-level truth. We introduced
\emph{Repository Supervision Auditing (RSA)}, a practical framework for auditing supervision before model
development: it validates repository labels against expert image-level annotations, identifies the causes
of disagreement, constructs balanced, quality-controlled cohorts, and evaluates downstream models under
deployment-oriented conditions. Applied to cardiomegaly in MIMIC-CXR, RSA revealed substantial
disagreement between repository-derived and expert image-level labels, driven primarily by reporting
practice rather than report-extraction errors.

As public imaging repositories increasingly underpin foundation models and clinically deployed AI systems,
RSA establishes supervision validation as a distinct stage of the medical AI development pipeline. By
validating repository-derived labels before model training, benchmarking, and deployment, RSA complements
advances in model architecture, optimization, and foundation-model pretraining, providing a practical
pathway toward more trustworthy medical imaging AI.

\begin{credits}
\subsubsection{\ackname}
The authors thank the SURA radiology team for their expert image-level annotations. This study used the MIMIC-CXR database, accessed under credentialed access through PhysioNet in accordance with the PhysioNet Credentialed Health Data Use Agreement (v1.5.0)~\cite{johnson2019mimic,goldberger2000physionet}; no attempt was made to re-identify any individual or institution. Computational resources were provided by Microsoft Azure Machine Learning. This research received no external funding.

\subsubsection{\discintname}
This work was carried out within SURA's institutional medical-imaging AI program; the SURA-affiliated authors are employees or trainees of SURA, and N. S. D'Souza is employed by IBM Research. The authors declare no other competing interests.
\end{credits}

\bibliographystyle{splncs04}
\bibliography{mybibliography.bib}
\end{document}